\documentclass[prl,preprint,showpacs,amsfonts,amsmath,floatfix]{revtex4}

\usepackage{graphicx}
\usepackage{color}

\begin{document}

%=======================================================================================
\title{Four-dimensional photonic lattices and discrete tesseract solitons}

\author{D.~Juki\'{c}}
\affiliation{Department of Physics, University of Zagreb, Bijeni\v cka c. 32, 10000 Zagreb, Croatia}

\author{H.~Buljan}
\email{hbuljan@phy.hr}
\affiliation{Department of Physics, University of Zagreb, Bijeni\v cka c. 32, 10000 Zagreb, Croatia}

\date{\today}

\begin{abstract}
We theoretically study discrete photonic lattices in more than three 
dimensions and point out that such systems can exist in continuous three-dimensional 
(3D) space. 
We study discrete diffraction in the linear regime, and predict the existence of 
four-dimensional (4D) {\em tesseract} solitons in nonlinear 4D periodic photonic lattices. 
Finally, we propose a design towards a potential realization of such periodic 
4D lattices in experiments. 
\end{abstract}

\maketitle
%\narrowtext
%\newpage

Dimensionality is one of the key attributes of physical systems which determines 
their properties. This particularly holds for photonic lattices - linear or nonlinear 
photonic structures conventionally considered 
in one- and two-dimensional geometry, where the flow of light 
exhibits a plethora of intriguing phenomena which hold great potential 
for applications \cite{Fleischer2005,Lederer2008}. 
Paradigmatic phenomena occurring in nonlinear photonic lattices 
are discrete solitons \cite{Fleischer2005,Lederer2008}. Their 
prediction in one-dimensional (1D) photonic lattices in 1988 \cite{Demetri1988} 
has awaited 10 years for the first experimental observation \cite{Eisenberg1998}. 
With the suggestion \cite{Nicos2002} and the experimental realization of optically 
induced photonic lattices, both in one \cite{Fleischer2003} and two \cite{JasonNat} 
spatial dimensions, novel excitations such as vortices \cite{FleischerVortex,NeshevVortex} 
(that cannot occur in 1D) were discovered and explored. 
One- and two- dimensional photonic lattices discussed here \cite{Fleischer2005,Lederer2008} 
are also referred to as waveguide arrays, because they are continuous along 
another dimension (second or third) along which the light propagates. 
Photonic crystals provide the opportunity to study versatile linear (e.g., see \cite{JoannopoulosBook}) and 
nonlinear phenomena \cite{MarinNatMat} in three spatial dimensions. In those systems 
one usually applies (continuous) Maxwell equations, but discrete lattice 
models may also be used when the so-called Tight-Binding Approximation (TBA) 
is applicable \cite{Lidorikis1998}. However, to the best of our knowledge photonic 
lattices in more than three dimensions were not yet considered. 
Here we theoretically study discrete photonic lattices in more than three 
dimensions. We point out that such systems can exist in continuous 3D space, 
that is, their experimental realization is not hindered due to the 
properties of our space. The properties of discrete diffraction and 
four-dimensional (4D) tesseract solitons are presented in 4D linear and nonlinear 
periodic discrete lattices. Finally, we propose a design towards a 
potential realization of such periodic 4D lattices in experiments.

The fact that a discrete lattice can have its dimension larger than the continuous space 
it is embedded in is known from complex networks \cite{Havlin2011}. 
For example, the Internet can be regarded as a network of dimension close to 4.5, 
and the network of airports close to 3, even though they are embedded on a 2D 
surface of Earth \cite{Havlin2011}. Complex networks have been scarcely considered in optical systems. 
We point out at an interesting concept of complex networks of interacting fields 
called solitonets \cite{Kaminer2008,Kaminer2010}, where the interaction dynamics at 
each individual node in the system has infinite degrees of freedom \cite{Kaminer2008,Kaminer2010}.

We begin with a paradigmatic model - the discrete nonlinear 
Schr\"odinger equation (DNLS) - which describes dynamics of light 
in lattices of various dimensions:
\begin{equation}
i \frac{d \psi_{\alpha}(t) }{ dt } = 
-J \sum_{\beta \in N_\alpha}  \psi_{\beta}
-\gamma |\psi_{\alpha}|^2 \psi_{\alpha},
\label{DNLSE}
\end{equation}
where $\psi_{\alpha}$ is the complex amplitude describing the field, 
$N_\alpha$ denotes the sites that are coupled to the site $\alpha$, 
$J$ is the coupling (hopping) parameter that we assume to be equal between 
all coupled sites, and $\gamma$ is the strength of the nonlinearity. 
Throughout the paper we use the following normalization: $\sum_\alpha |\psi_\alpha|^2=1$.
This model was succesfully used to describe dynamics of light in 
1D and 2D photonic lattices (waveguide arrays) \cite{Fleischer2005,Lederer2008}. 
It is applicable when the lattice wells are sufficiently deep, such that 
each well has a well defined resonance, and the coupling between different lattice 
sites is weak. Thus, one can think of this model as describing a system of 
weakly coupled high-Q resonators.

In theory, any two pairs of resonators can be coupled thus yielding versatile structures of complex 
networks of resonators, which calls for a more rigorous definition of dimension.
The dimensionality of any such network can be calculated by the 
following procedure: Let us choose one resonator and calculate the number  of resonators [call it $N(l)$]
that one can reach in $l$ or fewer connections. The number $N(l)$ scales as $N(l)\sim l^{D}$ when 
$l\rightarrow \infty$. This procedure is somewhat altered from that 
used in \cite{Havlin2011} for usual complex networks due to the fact that it is the 
possibility of coupling rather than the Euclidean 
distance between the resonators that matters here. It is straightforward to verify 
that the dimensionality of a simple cubic lattice corresponds to half the number of nearest 
neighbors, but we emphasize that this is not a generally valid prescription. To see that note that 
the well known body centered cubic lattice (BCC) or face centered (FCC) lattices have dimension three as expected.

Up to this point the theoretical model (\ref{DNLSE}) was general in a sense that 
coupling between any two pairs of resonators was possible. From this point on we focus on 
4D "simple cubic" lattices defined as follows: every resonator is labeled by 
four indices, $\alpha=(i,j,k,l)$, and it is coupled to eight resonators labeled 
by $(i+1,j,k,l)$, $(i-1,j,k,l)$, $(i,j+1,k,l)$, $(i,j-1,k,l)$, $(i,j,k+1,l)$, $(i,j,k-1,l)$, 
$(i,j,k,l+1)$, and $(i,j,k,l-1)$. We proceed with a discussion of light propagation 
phenomena encountered in these 4D "simple-cubic" lattices. 
The first question concerns diffraction of light, i.e., dynamics when nonlinearity is absent ($\gamma=0$). 
The phenomenon of discrete diffraction has been addressed many times in 1D and 2D systems
(e.g., see \cite{Eisenberg1998,JasonNat}), yielding a characteristic pattern of lobes 
spreading during propagation. 
In 4D it is impossible to visualize such a pattern and therefore to characterize diffraction 
we utilize the concept of inverse participation ratio, $I(t)=  1 / \sum_{\alpha} I(t)^2_{\alpha}$, 
where $I(\alpha)=|\psi_{\alpha}(t)|^2$ is the intensity of light.
A typical question that we wish to address is: if we excite a single 
resonator, how does the excitation spread through the system? 
For a $D$-dimensional "simple-cubic" lattice, propagation of the complex amplitude 
is given by 
\begin{equation}
\psi_{j_1 j_2\ldots j_D} = 
\prod_{\alpha=j_1 \ldots j_D}  i^{\alpha} {\mathcal J}_{\alpha}(2Jt),
\end{equation}
where the initially excited (at $t=0$) site is $\alpha_0=(0,\ldots,0)$ with amplitude 
$1$, and ${\mathcal J}_{n}$ is a Bessel function of order $n$. 
The evolution of inverse participation ratio (IPR) is asymptotically then  
\begin{equation}
I(t)= \left(\sum\limits_{n=-\infty}^{\infty} \mathcal{J}_n(2Jt)^4 \right)^{-D} 
\sim \left( \frac{2Jt}{\log{2Jt}} \right)^D,
\label{Ise}
\end{equation}
where we have utilized a formula from Ref. \cite{Martin2008} to express the sum. 
In what follows, for simplicity we take $J=1$.

\begin{figure}
\begin{center}
\mbox{\includegraphics[width=0.85\textwidth]{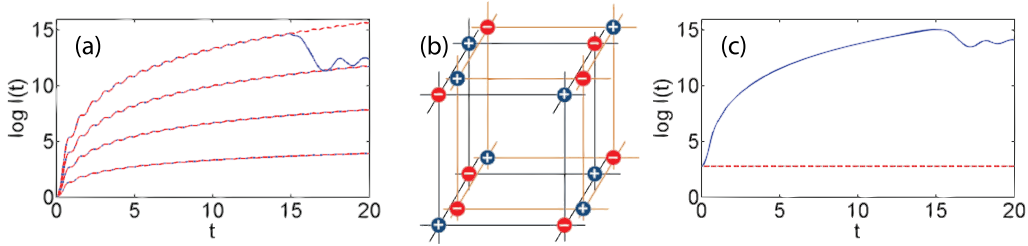}}
\caption{ \label{log_ipr}
(color online) Discrete diffraction in 4D lattices and tesseract solitons. 
(a) Evolution of the inverse participation ratio $I(t)$ during discrete diffraction 
for linear lattices with $D=1,2,3,4$ (bottom up). Red dashed lines depict $I(t)$ for an 
infinite lattice, whereas blue solid lines correspond to a finite size system with 
$N=2^{24}$ sites and periodic boundary conditions. For $D=4$ the boundary is reached 
at $t\approx 15$ and the IPR starts to oscillate. 
(b) A schematic illustration of a tesseract soliton. The phases at nearest neighbor 
nodes differ by $\pi$, $\gamma/J=630$. 
(c) Inverse participation ratio dynamics of a tesseract soliton with (black dotted line) 
and without (red dashed line) noise demonstrating stability (the two lines are 
on top of each other). When the nonlinearity is turned off, the 
tesseract initial condition diffracts (solid blue).  
}
\end{center}
\end{figure}

It is evident that the inverse participation ratio asymptotically increases as a power 
law with a slow logarithmic modulation, where the power-law exponent equals dimension. 
The logarithmic modulation would not have been present in continuous systems and 
its existence occurs because of the presence of the lobes. The effective number 
of sites excited by the light is smaller during diffraction in discrete systems 
in comparison to volume in continuous systems where the lobes are not present. 
Figure \ref{log_ipr}(a) illustrates diffraction IPR-dynamics for a finite amount of 
time in a "simple cubic" lattice in $1,2,3$, and $4D$. 
Superimposed on the results for an infinite lattice (red dashed lines), we also 
show dynamics for a finite size lattice (blue solid lines) with $N=2^{24}$ sites 
and periodic boundary conditions. 
This is important because finite size effects will occur in practical realizations of 4D lattices. 
For example, our 4D lattice has the length of only $2^6=64$ sites along one dimension and 
the effect is clearly visible for large enough times. 

Next we consider a 4D lattice of nonlinear resonators that we model with a DNLS Eq. (\ref{DNLSE}). 
A paradigmatic nonlinear phenomenon that occurs in nonlinear lattices are discrete or lattice solitons. 
Versatile types of 1D-3D discrete solitons were predicted and/or observed in optics including 
bright on-site solitons \cite{Demetri1988,Eisenberg1998,JasonNat}, 
staggered solitons \cite{Fleischer2003,Kivshar1993}, vortex solitons 
\cite{FleischerVortex,NeshevVortex}, and octopole solitons \cite{Kevrekidis2005}.
The simplest type of soliton that one can consider in 4D lattices is a 4D 
on-site bright soliton that is centered on a single site. 
We have found this soliton by self-consistently numerically 
solving the stationary DNLS equation with the focusing nonlinearity $\gamma>0$ (not shown). 
It occurs only above some threshold value of the nonlinearity (this also holds for 2D and 3D 
solitons \cite{Nicos2D,Kevrekidis2005}). The relevant parameter here 
is in fact the ratio $\gamma/J$, because DNLS can be scaled; this means that by 
reducing the coupling parameter $J$ the effective nonlinearity $\gamma/J$ can be made 
stronger. On-site bright solitons are also found in 1D-3D photonic systems, and 
one can expect that other types of 1D-3D soliton excitations will also 
exist in 4D systems.

Here we predict a novel type of soliton that occurs solely in 4D lattices: 
the {\em tesseract} solitons equally excite 16 sites of a 4D cube (i.e., tesseract), 
and therefore cannot exist in 1D-3D systems. 
By using the methods outlined in Ref. \cite{Kevrekidis2005}, 
we find that these solitons exist when the neighboring sites are $\pi$ 
out-of-phase as illustrated in Fig. \ref{log_ipr}(b).
The stability of these solitons was checked numerically; Fig. \ref{log_ipr}(c)
shows the inverse participation ratio dynamics of a stable tesseract soliton, 
and its discrete diffraction when propagated without nonlinearity present. 
Without the $\pi$ out-of-phase feature the intensity on the neighboring sites 
would not repel, and the excitation would collapse. 

\begin{figure}
\begin{center}
\includegraphics[width=0.5 \textwidth]{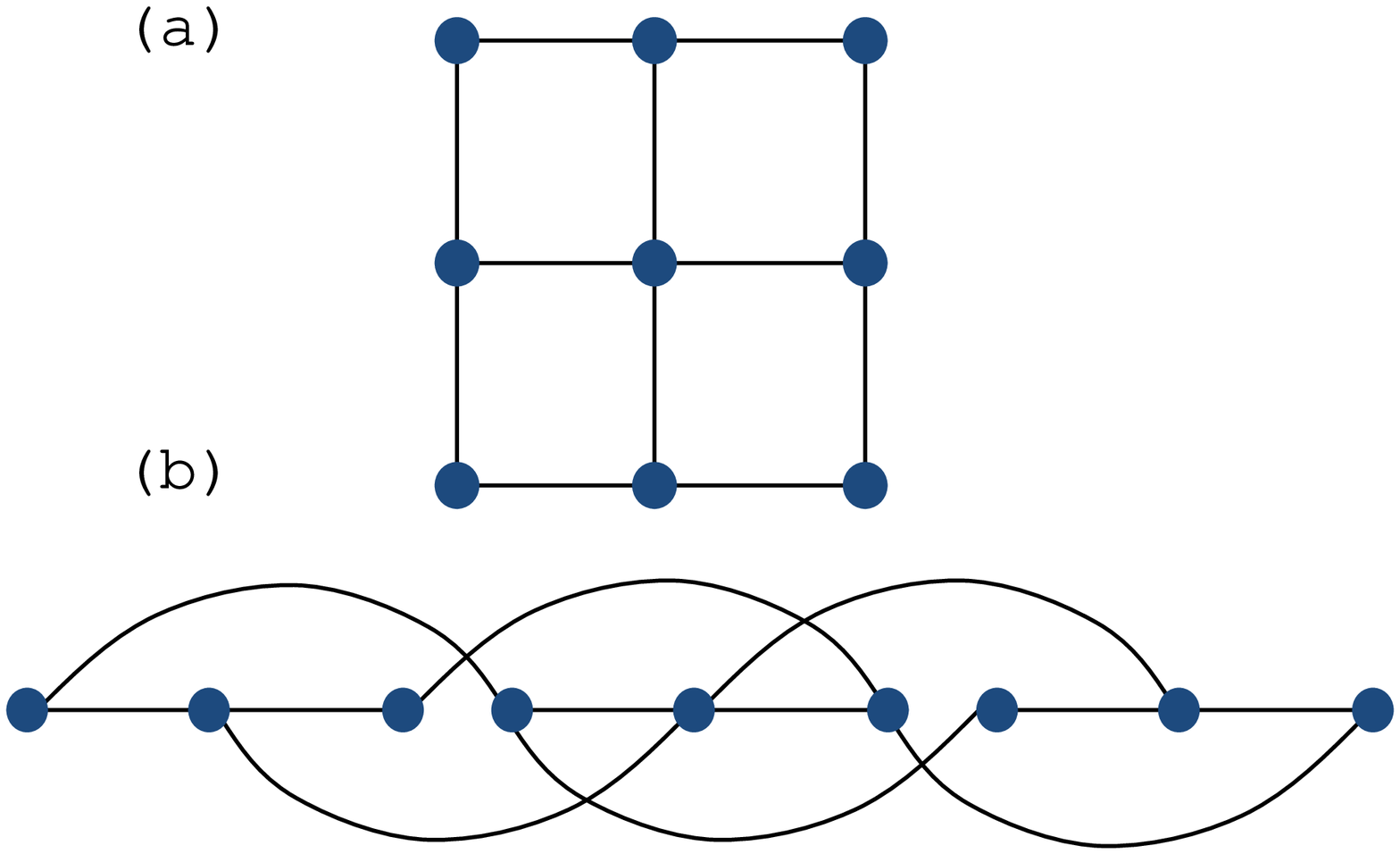}
\caption{ \label{idea}
(color online) Illustration of the idea for creating photonic lattices in more 
than three dimensions via complex networks of optical resonators. 
(a) A schematic illustration of a (finite size) two-dimensional lattice of 
coupled resonators. 
(b) The coupling scheme which is topologically fully equivalent to 
the one in (a), despite the fact that resonators are located on 
a 1D line. 
By the same token a 4D (discrete) lattice can be embedded in 
a continuous 3D space, see text for details. 
}
\end{center}
\end{figure}

Up to this point we have theoretically considered a model 
which can represent lattices in more than three dimensions, and analyzed 
some linear and nonlinear phenomena in such systems. However, one may 
say that these theoretical considerations are not more than academic curiosity 
because experiments are performed in 3D continuous space. We point out 
that discrete photonic lattices with dimensionality greater than three 
can exist in continuous 3D space. To illustrate that fact compare a 
system of coupled resonators schematically illustrated in Figs. \ref{idea}(a) and (b). 
(the coupled resonators are connected by lines). 
For the sake of the argument let us assume that the coupling parameters 
between all coupled resonators are equal, and zero otherwise. 
The configuration in Fig. \ref{idea}(a) is evidently a discrete 2D lattice. 
However, the system sketched in Fig. \ref{idea}(b) is fully equivalent to that of Fig. \ref{idea}(a). 
Thus, a 2D network of resonators can be constructed by embedding these resonators 
in 1D geometry (on a straight line), provided that connections between distant 
(in Euclidian sense) resonators can be made. By using this line of reasoning it is evident that 
the existence of discrete photonic lattices of dimension larger than three, describable with 
model (\ref{DNLSE}), is not hindered by the dimensionality of our space.

The limitations on the dimensionality and structure of experimentally realizable 
discrete lattices depend on our ability to construct coupled resonators that 
may be on distant locations. This is not an easy task because usually 
high-Q resonators are coupled via evanescent coupling, which implies that 
they have to be close to each other because the light is tightly bound to 
the resonators and evanescent fields extend only a few wavelengths. 
However, in a recent work Sato {\em et al.} have demonstrated that two distant photonic high-$Q$ cavities can be 
coupled by using an appropriate waveguide with mirrors at its ends \cite{Sato2012}.
The realization of the cavities and coupler was made in a photonic crystal 
structure \cite{Sato2012}. 
Importantly, many Rabi oscillations were observed in that system \cite{Sato2012} which 
provides a great promise for the future construction of 4D lattices.

\begin{figure}
\begin{center}
\includegraphics[width=0.5\textwidth]{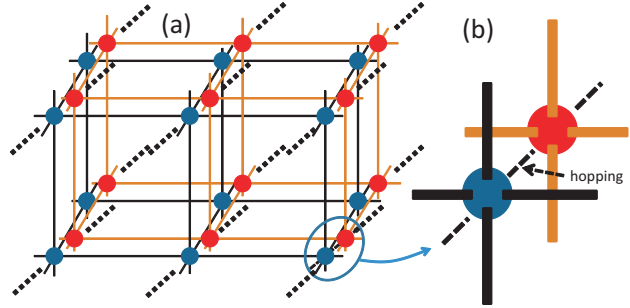}
\caption{ \label{4Dlatt}
(color online) Proposal for the design of a 4D photonic lattice. 
(a) Illustration of a "simple cubic" 4D lattice.
Waveguides enable coupling between distant resonators along the edges of 
3D cubes. Evanescent coupling (hopping) along the diagonal enables construction 
of the 4th dimension for the discrete lattices, see encircled region enlarged in (d). 
For clarity only two sublattices are displayed in (a); the number of sublattices 
that can be added along the 4th dimension depends on the size of each individual cavity 
$R$, and the lattice size $a_l$ of every 3D sublattice. See text for details. 
}
\end{center}
\end{figure}

Two ingredients towards the realization of more than three-dimensional discrete photonic 
lattices are thus present: (i) the fact that more than three dimensional photonic lattices can in principle 
exist in our continuous 3D space, and (ii) distant high-Q cavities can be 
coupled as was demonstrated in Ref. \cite{Sato2012}. 
A proposal of a completely specified photonic structure which would lead to 
an experimental realization of more than 3D photonic lattices is beyond the 
scope of the present work. However, here we propose a design which 
seems as a viable path towards the realization of 4D "simple-cubic" photonic lattices. 
It is based on a combination of waveguide coupling and evanescent coupling between 
resonators. The design is illustrated in Fig. \ref{4Dlatt}. First consider a 3D 
simple-cubic 3D lattice of cavities coupled with waveguides.
Let the unit vectors of this 3D simple-cubic lattice be ${\mathbf a_1}$, ${\mathbf a_2}$, and 
${\mathbf a_3}$. The distance between two adjacent sites 
$a_l$ should be much larger than the size of an individual cavity $R$. 
Consider now that we place two such simple cubic structures next to each 
other such that each site of the second lattice is displaced by the vector 
${\mathbf d}=d ({\mathbf a_1}+{\mathbf a_2}+{\mathbf a_3})/\sqrt{3}$ from the first one,
where $d$ is slightly larger than $2R$, so that evanescent coupling (i.e. hopping) 
from one sublattice to the other is possible [see Figs. \ref{4Dlatt}(a) and (b)].
By adjusting the distances $d$ and $a_l\gg R$, the coupling constant 
between two sites of the same sublattice (coupled via waveguides) and 
the hopping parameter between two adjacent sublattices, can in principle be 
made approximately equal. 
By this procedure we have constructed a bi-cubic lattice. However, 
since $a_l\gg d\sim 2R$, we can add a number (roughly up to $a_l\sqrt{3}/d$) of 
sublattices along the direction ${\mathbf d}$ thereby creating a finite size 4D lattice. 
If $a_l$ is a bit more than 10 times larger than $d$, then a finite size 4D simple cubic 
lattice with $10^4$ sites can be constructed. 
For resonators made of nonlinear materials this construction 
yields 4D nonlinear photonic lattices.

Let us comment on the finite size effects and scaling associated 
with realizations of these "simple-cubic" 4D lattices. From the design scheme presented in Fig. \ref{4Dlatt} 
we see that along the fourth discrete dimension the photonic lattice would have to be finite; 
its size along this direction, i.e., the number of sublattices that one can add,  
depends on the ratio of the cube edges $a_l$ and the distance $d$ between the resonators 
along the fourth discrete dimension. Thus, to build a larger 4D lattice one needs larger $a_l$. 
In practice all lattices are finite (including 1D-3D) and if they are sufficiently large 
the finite size effects would be unimportant. Our calculations show that 
4D lattices with $10^4$ sites (10 sites along one dimension) can already exhibit 
4D behavior for some phenomena like solitons.

In conclusion we have studied discrete photonic lattices in more than three dimensions, 
and pointed out that such systems can exist in continuous 3D space. 
We have studied discrete diffraction in the linear regime, and predicted the existence of 
4D {\em tesseract} solitons in nonlinear 4D periodic photonic lattices.
These novel structures would open the way 
for investigating new optical phenomena that one does not encounter in 
usual 1D-3D systems, but could also provide us with better understanding of 
dimensionalities beyond 3D which is of fundamental importance. 
We envision the study of versatile novel types of solitons and instabilities 
in these systems including vortex like structures, gap solitons, surface states 
and surface solitons, incoherent light dynamics, and studies of 
quantum optical phenomena. These structures could also yield schemes 
and opportunities for creating novel optical devices.

{\em Note added.} When this work was in its final stages we became aware of a 
paper \cite{Lewenstein2012} which proposes realization of a 4D quantum models 
by using ultracold atoms in optical traps, where the 4th dimension is encoded 
in the internal states of the atoms providing an extra degree of freedom.

This work is supported by the Croatian Ministry of Science (Grant No. 119-0000000-1015).

%==============================================================================


\begin{thebibliography}{99}
%%%%%%%%%%%%%%%%%%%%%%%%%%%%%%%%%%%%%%%%%%%%%%%%%%%%%%%%%%%%%%%%%%%%%%%%%%%%%%%

\bibitem{Fleischer2005}
J.W. Fleischer, G. Bartal, O. Cohen, T. Schwartz, O. Manela, B. Freedman, 
M. Segev, H. Buljan, N.K. Efremidis, 
Opt. Express {\bf 13}, 1783 (2005).

\bibitem{Lederer2008}
F. Lederer, G.I. Stegeman, D.N. Christodoulides, G. Assanto, M. Segev, 
Y. Silberberg, 
Phys. Reports {\bf 463}, 1 (2008).

\bibitem{Demetri1988}
D.N. Christodoulides and R.I. Joseph, 
Opt. Lett. {\bf 13}, 794 (1988). 

\bibitem{Eisenberg1998}
H.S. Eisenberg, Y. Silberberg, R. Morandotti, A.R. Boyd, and J.S. Aitchison, 
%"Discrete spatial optical solitons in waveguide arrays," 
Phys. Rev. Lett. {\bf 81}, 3383 (1998).

\bibitem{Nicos2002}
N.K. Efremidis, S. Sears, D.N. Christodoulides, J.W. Fleischer, and M. Segev, 
%"Discrete solitons in photorefractive optically induced photonic lattices," 
Phys. Rev. E {\bf 66}, 046602 (2002).

\bibitem{Fleischer2003}
J.W. Fleischer, T. Carmon, M. Segev, N.K. Efremidis, and D.N. Christodoulides, 
%"Observation of discrete solitons in optically induced real time waveguide arrays," 
Phys. Rev. Lett. {\bf 90}, 023902 (2003).

\bibitem{JasonNat}
J.W. Fleischer, M. Segev, N.K. Efremidis, and D.N. Christodoulides, 
%"Observation of two-dimensional discrete solitons in optically induced nonlinear photonic lattices," 
Nature {\bf 422}, 147 (2003).

\bibitem{FleischerVortex}
J.W. Fleischer, G. Bartal, O. Cohen, O. Manela, M. Segev, J. Hudock, and 
D.N. Christodoulides, 
%"Observation of vortex-ring 'discrete' solitons in 2D photonic lattices," 
Phys. Rev. Lett. 92, 123904 (2004).

\bibitem{NeshevVortex}
D.N. Neshev, T.J. Alexander, E.A. Ostrovskaya, Y.S. Kivshar, H. Martin, I. Makasyuk, and Z. Chen,
%"Observation of discrete vortex solitons in optically induced photonic lattices," 
Phys. Rev. Lett. {\bf 92}, 123903 (2004).

\bibitem{JoannopoulosBook}
J.D. Joannopoulos, S.G. Johnson, J.N. Winn, and R.D. Meade,
{\em Photonic Crystals: Molding the Flow of Light}
(Princeton University Press, Princeton, 2008).

\bibitem{MarinNatMat}
M. Solja\v{c}i\'{c} and J.D. Joannopoulos, 
Nat. Materials {\bf 3} 211, (2004).

\bibitem{Lidorikis1998}
E. Lidorikis, M.M. Sigalas, E.N. Economou, and C.M. Soukoulis
Phys. Rev. Lett. {\bf 81}, 1405 (1998).

\bibitem{Havlin2011}
L. Daqing, K. Kosmidis, A. Bunde, and S. Havlin,
Nat. Physics {\bf 7}, 481 (2011). 

\bibitem{Kaminer2008}
I. Kaminer, M. Segev, A.M. Bruckstein, and Y.C. Eldar, 
Proc. R. Soc. A {\bf 465}, 1093 (2009). 

\bibitem{Kaminer2010}
I. Kaminer, M. Segev, A.M. Bruckstein, 
Phys. Rev. Lett. {\bf 105}, 083901 (2010).

\bibitem{Sato2012}
Y. Sato, Y. Tanaka, J. Upham, Y. Takahashi, T. Asano, and S. Noda,
Nat. Photonics {\bf 6}, 56 (2012). 

\bibitem{Martin2008}
P.A. Martin, 
J. Phys. A.: Math. Ther. {\bf 41}, 015207 (2008).

\bibitem{Kivshar1993}
Y.S. Kivshar, 
%"Self-localization in arrays of defocusing waveguides," 
Opt. Lett. 18, 1147 (1993).

\bibitem{Kevrekidis2005}
R. Carretero-Gonzalez, P.G. Kevrekidis, B.A. Malomed, and D.J. Frantzeskakis,
Phys. Rev. Lett. {\bf 94}, 203901 (2005).

\bibitem{Nicos2D}
N.K. Efremidis, J. Hudock, D.N. Christodoulides, J.W. Fleischer, O. Cohen, and M. Segev, 
%"Twodimensional optical lattice solitons," 
Phys. Rev. Lett. {\bf 91}, 213906 (2003).

\bibitem{Lewenstein2012}
O. Boada, A. Celi, J.I. Latorre, and M. Lewenstein,
Phys. Rev. Lett. {\bf 108}, 133001 (2012). 

\end{thebibliography}
\end{document}